## METHODOLOGY

**Open Access**

# A GPU based multidimensional amplitude analysis to search for tetraquark candidates

Nairit Sur[1]*[†] , Leonardo Cristella[2†], Adriano Di Florio[2†] and Vincenzo Mastrapasqua[2†]

*Correspondence: nairit.sur@cern.ch
[†]Nairit Sur, Leonardo Cristella, Adriano Di Florio and Vincenzo Mastrapasqua contributed equally to this work
[1] Tata Institute of Fundamental Research, Mumbai, India
Full list of author information is available at the end of the article

**Abstract**

The demand for computational resources is steadily increasing in experimental high energy physics as the current collider experiments continue to accumulate huge amounts of data and physicists indulge in more complex and ambitious analysis strategies. This is especially true in the fields of hadron spectroscopy and flavour physics where the analyses often depend on complex multidimensional unbinned maximum-likelihood fits, with several dozens of free parameters, with an aim to study the internal structure of hadrons. Graphics processing units (GPUs) represent one of the most sophisticated and versatile parallel computing architectures that are becoming popular toolkits for high energy physicists to meet their computational demands. GooFit is an upcoming open-source tool interfacing ROOT/RooFit to the CUDA platform on NVIDIA GPUs that acts as a bridge between the MINUIT minimization algorithm and a parallel processor, allowing probability density functions to be estimated on multiple cores simultaneously. In this article, a full-fledged amplitude analysis framework developed using GooFit is tested for its speed and reliability. The four-dimensional fitter framework, one of the firsts of its kind to be built on GooFit, is geared towards the search for exotic tetraquark states in the $B^0 \to J/\psi K \pi$ decays and can also be seamlessly adapted for other similar analyses. The GooFit fitter, running on GPUs, shows a remarkable improvement in the computing speed compared to a ROOT/RooFit implementation of the same analysis running on multi-core CPU clusters. Furthermore, it shows sensitivity to components with small contributions to the overall fit. It has the potential to be a powerful tool for sensitive and computationally intensive physics analyses.

## Introduction

Big data usually includes data sets with sizes beyond the ability of commonly used software tools to capture, curate, manage, and process data within a tolerable elapsed time, thus calling for parallel computing tools to analyze them [1, 2]. Modern particle accelerators such as the Large Hadron Collider (LHC) [3] at European Organization for Nuclear Research (CERN) produce data at a phenomenal rate [4]. CERN operates the largest particle physics laboratory in the world providing requisite infrastructures for research in high-energy physics, viz. powerful computing facilities primarily used to store and analyse data from experiments, as well as to simulate events. The LHC is a superconducting accelerator and collider of protons and heavy ions at teraelectronvolt (TeV) energy scales. An electronvolt is a unit of energy defined as the amount of kinetic energy gained or lost by a single electron accelerating from rest through an electric potential difference of one volt in vacuum. The LHC consists of a 27-km circular underground tunnel of superconducting magnets with a number of accelerating structures to boost the energy of the particles along the way. The accelerated protons flow in opposite directions through two parallel beam pipes of the circular LHC tunnel and collide with each other at four points where the beam pipes cross each other. Massive and intricate detectors, such as ATLAS [5], CMS [6], and LHCb [7], are built around these collision points to detect the huge number of particles created due to the 600 million collisions taking place per second. Such particle accelerators and the associated detectors are collectively referred to as "collider experiments". The LHC experiments represent about 150 million sensors delivering data at the rate of 40 MHz. The raw data flow from the LHC detectors exceeds 500 exabytes per day which is almost 200 times more than all the other sources combined in the world. Even after preserving only a fraction of that data stream for physics analysis, hundreds of petabytes of complex data are stored and processed [8, 9].

One of the key challenges in analysing and interpreting these data is to accurately model the distributions of observable quantities in terms of the physics parameters of interest. The result of repeating an experiment (like tossing a coin or rolling dice) many times does not lead to the same result but produce a distribution of answers. The form of the distribution depends on the nature of the experiment and can be represented by mathematical models [10]. In addition, a large number of other parameters may be needed to accurately describe the resolution and efficiency of complex detectors. These mathematical models are constructed in terms of probability density functions (PDFs) [11] normalized over the allowed range of observables with respect to the parameters. Due to the large amount of data, as well as the ever-increasing complexity of physics models, the running time of estimation of the parameters ("fitting" [10]) has become a major bottleneck.

The PDFs become particularly complex while probing different aspects of quantum chromodynamics (QCD), the quantum field theoretical description of strong interaction between the quarks and gluons [12]. Quarks and gluons are elementary constituents of matter. They combine together to form composite particles called hadrons. The most common hadrons, namely protons and neutrons, form atomic nuclei and are thus responsible for most of the mass of the visible matter in the universe. Hadrons are generally of two types—baryons (bound state of three quarks) and mesons (bound states of a quark and an antiquark). The study of masses and decays of hadrons is called hadron spectroscopy which is a key to understand QCD. Due to the complex nature of this non-abelian gauge theory including peculiar features like "colour confinement" and "asymptotic freedom" [13–15], it is very hard to study the nature of this interaction analytically, especially at low energy regimes. In the last 15 years, experimental evidence has been mounting [16] for a large number of multiquark bound states that are allowed in principle by QCD but do not fit the expectations for the conventional quark model (i.e., the baryons or the mesons) and relative spectra. These new particles are often called "exotic" states. The exact nature of many of these states still remains a puzzle; even though some of them are confirmed by multiple experiments, not all the quantum numbers of these states have yet been determined. Spectroscopic studies of such heavy-flavor states can provide a deeper understanding of the underlying dynamics of quarks and gluons at the hadron mass scales as well as a valuable insight into various QCD inspired phenomenological models [16, 17].

The charged charmonium-like $Z$ states, which are strong candidates for tetraquark states with a possible quark content of $|c\bar{c}d\bar{u}\rangle$, can be studied in ongoing collider experiments, ATLAS, Belle II [18], BESIII [19], CMS, and LHCb. To ascertain, with a high degree of statistical significance, the presence of such intermediate states in three-body decays $B^0 \to \psi(nS)K^+\pi^-$, complex multidimensional unbinned maximum-likelihood (UML) [11] fits on tens of thousands of data points, with several dozens of free parameters, must be performed, thus requiring a considerable amount of computational resources. The traditional high-energy physics (HEP) analysis tools such as ROOT [20] and RooFit [21], which are designed to run on CPUs, require excessively long processing times amounting to days even when they are run on servers comprising several multi-core CPUs.

In this article, we explore the scope of an advanced GPU-accelerated computing framework to reduce the processing times of such complex multidimensional fits frequently occurring in the field of HEP. We expand the usability of existing software keeping in mind the particular needs of a typical HEP analysis. This article starts with a comprehensive overview of existing literature in the emerging field of GPU-assisted HEP analysis, followed by a detailed methodology of a four-dimensional amplitude analysis. The findings are discussed in "Results" section and concluding remarks are elaborated in "Discussion" and "Conclusion" sections.

Our framework is based on the novel GPU based GooFit [22, 23] package. GooFit is an open-source analysis tool, presently under development, which can be used in the HEP applications for parameters estimation, and which interfaces ROOT to the CUDA parallel computing platform on NVIDIA GPUs [24]. GPU-accelerated computing enhances application performances by offloading a sequence of elementary but computationally intensive operations to the GPU to be processed in parallel, while the remaining code still runs on the CPUs. MINUIT [25] is a numerical minimization program that searches for a minimum in a user-defined function with respect to one or more parameters using several different methods as specified by the user. MINUIT cannot be distributed as an executable binary to be run by a relatively unskilled user. The user must write and compile a subroutine defining the function to be optimized, and oversee the optimization process. GooFit acts as an interface between MINUIT and the GPU, which allows any PDF to be evaluated in parallel over a huge amount of data. Fit parameters are estimated at each negative-log-likelihood (NLL) minimization step on the *host side* (CPU) while the PDF/NLL is evaluated on the *device side* (GPU). GooFit is still a limited open-source tool, being mainly developed by the users themselves for their specific needs. A very few applications in HEP analysis have been designed using GooFit. Significant sections needed for our fit implementation have been either newly encoded or adapted starting from the existing classes and methods.

## State-of-the-art literature

The need for GPU-based analysis frameworks to meet the demands of current and future HEP experiments has been acknowledged within the community for quite some time [26]. To that end, the GPU-based GooFit package was developed to mimic the functionalities and flexibilities of the widely popular RooFit one.

GooFit is designed to minimize the amount of CUDA coding required by a general user while exploiting the full potential of GPU parallelization. GooFit objects, viz. PDFs, can be created and combined in standard C++ if the PDFs are already encoded in existing classes. However, the available classes are limited in number and many other functionalities widely used for HEP analyses are not yet developed within the framework. The general algorithm to develop new PDF models within GooFit and test their functionalities involves coding with the help of CUDA while keeping in mind the complex data organization in GooFit that facilitates an efficient transfer of bytes between the host and device. During the fitting process the PDF must be normalized accurately. As it is not feasible to find an analytic expression for complicated functions in general, the normalization is computed numerically which requires evaluation of the function at several million phase space points.

One of the first performance comparison studies of GooFit vs. RooFit was conducted in Ref. [27]. Here, a high-statistics toy Monte Carlo technique was implemented for a simple 2D PDF model with a few parameters and the fit performances were compared for binned maximum likelihood fits. A further extension can be found in Ref. [28] where pseudo-experiments are coupled with a complex clustering technique in order to include the Look-Elsewhere-Effect when assessing the statistical significance of a new physics signal.

Models of higher complexity viz. time-dependent Dalitz plot analysis and model-independent partial wave analysis have been gradually added to the GooFit package as demonstrated in Refs. [29–31]. All these are extensions of the standard UML fit of the Dalitz plot [32], in which the matrix element describing the decay process is represented by a coherent sum of quantum mechanical amplitudes.

The models developed so far within GooFit could not perform a full-fledged amplitude analysis fit for complex processes such as a pseudoscalar meson decaying into at least one vector state along with another zero- or a higher-spin particle, with an eventual four-particle final state. The complexity arises due to an additional angle-dependent part of the PDF needed to describe the more complicated decay dynamics. This functionality has now been introduced for the first time and is described in detail in this article.

## Methodology

### An amplitude analysis of the three body decay $B^0 \to J/\psi K \pi$

The rare exotic $Z$ states can appear as $J/\psi \pi$ resonances in the quasi two-body decay $B^0 \to Z^- K^+ \to J/\psi \pi^- K^+$, where the $J/\psi$ decays into a $\mu^+ \mu^-$ pair (inclusion of the charge conjugate mode $\bar{B}^0 \to Z^+ K^- \to J/\psi \pi^+ K^-$ is always implied). However, the decay process is dominated by the intermediate $K^*(\to K\pi)$ resonances in the quasi two-body decay $B^0 \to J/\psi K^*$ [33]. These ten kinematically allowed kaonic resonances can interfere with one another as well as with the $Z$ states.

 

Three-body decays with intermediate resonant states, such as $P \to D_1 + D_{\text{res}}$, $D_{\text{res}} \to D_2 + D_3$, are generally analysed using a technique pioneered by Dalitz [32]. Here, $P$ is the parent particle, $D_1$ is one of its daughters, $D_{\text{res}}$ is the other daughter which, being an intermediate resonance, decays into $D_2$ and $D_3$. A two-dimensional scatter plot of $m^2_{D_1 D_2}$ vs. $m^2_{D_2 D_3}$ (invariant mass squared of any two daughters), known as the Dalitz plot, shows a nonuniform distribution due to the interfering intermediate resonances, thus to the decay dynamics. If at least one of the three daughters in the decay is a vector state instead of being a pseudoscalar, the traditional Dalitz plot approach becomes insufficient as the angular variables are implicitly integrated over, leading to a loss of information about angular correlations among the decay products.

### The K*-only model

The kinematics of the process $B^0 \to J/\psi K \pi$, $J/\psi \to \mu^+ \mu^-$ can be completely described by a four-dimensional variable space:

$$\Phi \equiv \left(m_{K\pi}, m_{J/\psi \pi}, \theta_{J/\psi}, \varphi\right). \tag{1}$$

The two angles, $\theta_{J/\psi}$ and $\varphi$ are illustrated in Fig. 1. The number of dimensions required to describe any decay process is given by the difference between the degrees of freedom of the system and the total number of constraints. A three-body decay in general has twelve degrees of freedom due to the four-momenta of each particle. As one of the particles in the $B^0 \to J/\psi K \pi$ decay is a vector state (spin 1), it has two extra degrees of freedom. The corresponding constraints are the conservation of four-momenta, the three masses, and the three euler angles. Thus the number of dimensions required becomes $(12 + 2 - 4 - 3 - 3) = 4$.

The relativistic Breit–Wigner (BW) function is a continuous probability distribution used to model resonances (unstable particles). The total decay amplitude of $B^0 \to J/\psi K \pi$ is represented by a coherent sum of the BW contributions associated with all the kinematically allowed intermediate resonant states. Simple field theory assumes all particles to be point like. In real life, however, the finite size of bound

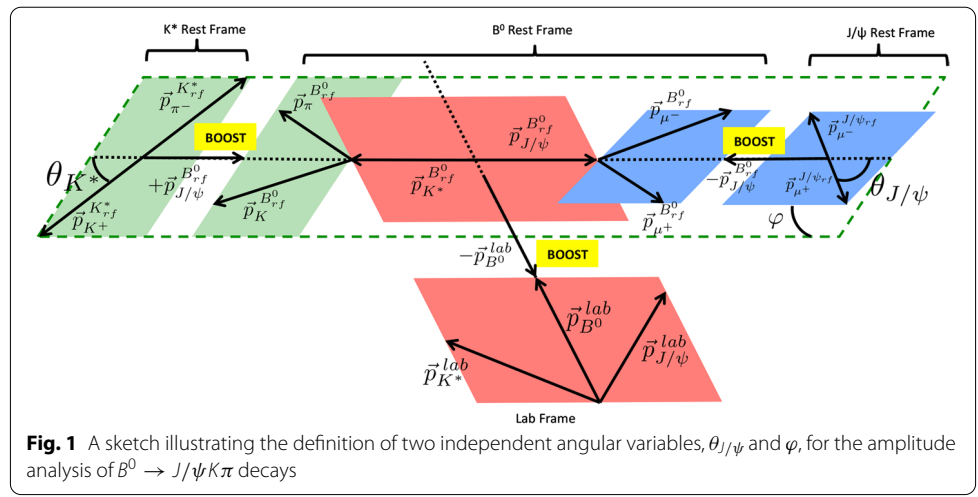

**Fig. 1** A sketch illustrating the definition of two independent angular variables, $\theta_{J/\psi}$ and $\varphi$, for the amplitude analysis of $B^0 \to J/\psi K \pi$ decays

states of hadrons is modeled by form factors that are used to modify the original BW shape. The angle-independent part of the decay amplitude for each resonance $R$ is given by [34]:

$$A^R\left(m_R^2\right) = \frac{F_B^{(L_B)}\left(\frac{p_B}{M_B}\right)^{L_B} F_R^{(L_R)}\left(\frac{p_R}{m_R}\right)^{L_R}}{M_R^2 - m_R^2 - iM_R\Gamma(m_R)}, \quad (2)$$

where the mass-dependent width of $R$ is:

$$\Gamma(m_R) = \Gamma_0 \left(\frac{p_R}{p_{R_0}}\right)^{2L_R+1} \left(\frac{M_R}{m_R}\right) F_R^2, \quad (3)$$

and

- $m_R$ is the running invariant mass of the two daughters of $R$ (e.g., $m_R = m_{K\pi}$ for a $K^*$);
- $M_B$ is the $B^0$ meson mass;
- $M_R$ is the nominal mass of $R$;
- $L_B$ ($L_R$) is the orbital angular momentum in the $B^0$ ($R$) decay;
- $p_B$ is the $B^0$ daughter momentum (i.e., $R$ momentum) in the $B^0$ rest frame;
- $F_B^{(L_B)}$ and $F_R^{(L_R)}$ are the Blatt–Weisskopf form factors [35] for $B^0$ and $R$ decay, respectively, with the superscript denoting the orbital angular momentum of the (sub-)decay;
- $\Gamma_0$ is the nominal width of $R$;
- $p_R$ and $p_{R_0}$ are the momenta of $R$ daughters in the former's rest frame, calculated from the running and pole mass of $R$, respectively.

For $K^*$ resonances with spin ($J$) of one or more units, $L_B$ can take several values ($S$, $P$, and $D$-waves for $J = 1$; $P$, $D$, and $F$-waves for $J = 2$; and $D$, $F$, and $G$-waves for $J = 3$). The lowest $L_B$ is taken as the default value while the other possibilities are considered as part of the uncertainty in measurements due to their small contributions.

A sequential decay of the $B^0$ meson via an intermediate resonance into a four-body final state involves multiple decay planes requiring the application of Lorentz boosts and rotations to go from one rest frame to another, as can be seen in Fig. 1. As the helicity remains invariant under both Lorentz boost and rotation, the angle-dependent part of the amplitude is obtained using the helicity formalism [36]. For each $K^*$ resonance, it is given by:

$$A_{\lambda\xi}^{K^*}(\Phi) = H_\lambda^{K^*} A^{K^*}\left(m_{K\pi}^2\right) d_{\lambda 0}^{J(K^*)}(\theta_{K^*}) e^{i\lambda\varphi} d_{\lambda\xi}^1(\theta_{J/\psi}), \quad (4)$$

where $A^{K^*}(m_{K\pi}^2)$, defined in Eq. (2), is explicitly written for $R \equiv K^*$ and

- $J(K^*)$ is the spin of the considered $K^*$ resonance;
- $\lambda$ is the helicity of the $J/\psi$ (the quantisation axis being parallel to the $K^*$ momentum in the $J/\psi$ rest frame). In general, $\lambda$ can take the values $-1$, 0 and 1. For $K^*$s with zero spin, only $\lambda = 0$ is allowed;

- $\xi$ is the helicity of the $\mu^+\mu^-$ system;
- $H_\lambda^{K^*}$ is the complex helicity amplitude for the decay via the intermediate $K^*$;
- $d_{\lambda 0}^{J(K^*)}(\theta_{K^*})$ and $d_{\lambda\xi}^1(\theta_{J/\psi})$ are the Wigner small-d functions that represent rotations;
- $\theta_{K^*}$ is the $K^*$ helicity angle, i.e. the angle between $K$ momentum in the $K^*$ rest frame and the $K^*$ momentum in the $B^0$ rest frame (Fig. 1);
- $\theta_{J/\psi}$ is the $J/\psi$ helicity angle, i.e. the angle between $\mu^+$ momentum in the $J/\psi$ rest frame and the $J/\psi$ momentum in the $B^0$ rest frame; and
- $\varphi$ is the angle between the $J/\psi \to \mu^+\mu^-$ and $K^* \to K\pi$ decay planes.

The signal density function, to be used in the UML fit, is obtained after appropriately summing over the helicity states and is given by:

$$S(\Phi) = \sum_{\xi=1,-1} \left| \sum_{K^*} \sum_{\lambda=-1,0,1} A_{\lambda\xi}^{K^*} \right|^2 \tag{5}$$

The sum over $K^*$ includes all kinematically allowed resonance states up to $m_{K\pi} = 2.183\,\text{GeV}$, namely $K_0^*(800)$, $K^*(892)$, $K^*(1410)$, $K_0^*(1430)$, $K_2^*(1430)$, $K^*(1680)$, $K_3^*(1780)$, $K_0^*(1950)$, $K_2^*(1980)$, and $K_4^*(2045)$. As the expression in Eq. (5) is sensitive only to the relative phases and amplitudes, we have the freedom to fix one overall phase and amplitude in the fit. The helicity amplitude of the $K^*(892)$, the dominant resonance, is chosen to be fixed, for $\lambda = 0$:

$$\left| H_0^{K^*(892)} \right| = 1, \quad \arg\left( H_0^{K^*(892)} \right) = 0. \tag{6}$$

The masses and widths of all the resonances are fixed to their world-average values [37].

### The LASS parametrization

Generally, *P*- and *D*-wave states are considered to be well described by narrow resonance approximations. For the $K\pi$ system, the low mass *S*-wave $K_0^*(800)$ appears as a broad peak calling for a more careful treatment. The LASS experiment at SLAC used an effective range expansion to model the low-energy behaviour of such $K\pi$ *S*-wave [38]. We use a similar parametrization where the angle-independent part of the amplitude is a nonresonant contribution interfering with the scalar $K_0^*(1430)$ BW amplitude:

$$A_{\text{LASS}} = \frac{m_{K\pi}}{q_{K\pi}} \sin\theta_B e^{i\theta_B} + 2e^{2i\theta_B} \frac{\left(m_{K_0^*(1430)}^2/q_{K_0^*(1430)}\right)\Gamma_{K_0^*(1430)}}{M_{K_0^*(1430)}^2 - m_{K\pi}^2 - iM_{K_0^*(1430)}\Gamma(m_{K\pi})}, \tag{7}$$

with

$$\cot\theta_B = \frac{1}{a\,q_{K\pi}} + \frac{1}{2}b\,q_{K\pi} \quad \text{and,} \quad a = 1.95\,\text{GeV}^{-1}, \quad b = 1.76\,\text{GeV}^{-1}, \tag{8}$$

where

- $m_{K\pi}$ is the running mass of the $K\pi$ system;
- $q_{K\pi}$ is the momentum of one of the $K^*$ daughters in the $K^*$ rest frame;

- $\Gamma(m_{K\pi})$ is the running resonance width.

Therefore, the signal density with the LASS parametrization for the low-mass $K\pi$ S-wave becomes,

$$S(\Phi) = \sum_{\xi=1,-1} \left| H_0^{\text{LASS}} A_{0\xi}^{\text{LASS}} + \sum_{K^{*\prime}} \sum_{\lambda=-1,0,1} A_{\lambda\xi}^{K^{*\prime}} \right|^2. \quad (9)$$

*Model including exotic Z resonances*

For the decay $B^0 \to KZ(\to J/\psi\pi)$, $J/\psi \to \mu^+\mu^-$ where the Z can either be a Z(4200), and/or a Z(4430), or any other exotic (charmonium-like) state, the angle-dependent amplitude is given as:

$$A_{\lambda'\xi}^Z(\Phi) = H_{\lambda'}^Z A^Z\left(m_{J/\psi\pi^+}^2\right) d_{0\lambda'}^{J(Z)}(\theta_Z) e^{i\lambda'\tilde{\varphi}} d_{\lambda'\xi}^1(\tilde{\theta}_{J/\psi}) e^{i\xi\alpha}, \quad (10)$$

where

- $J(Z)$ is the spin of the Z resonance, we consider only $1^+$ spin-parity of the Zs as per Belle's result [33];
- $\lambda'$ is the helicity of the $J/\psi$ (quantisation axis parallel to the $\pi$ momentum in the $J/\psi$ rest frame);
- $\xi$ is the helicity of the $\mu^+\mu^-$ system;
- $H_{\lambda'}^Z$ is the complex helicity amplitude for the decay via the intermediate Z;
- $d_{0\lambda'}^{J(Z)}(\theta_Z)$ and $d_{\lambda'\xi}^1(\tilde{\theta}_{J/\psi})$ are the Wigner small-d functions;
- $\theta_Z$ is the Z helicity angle, i.e. the angle between K and $\pi$ momenta in the Z rest frame;
- $\tilde{\theta}_{J/\psi}$ is the $J/\psi$ helicity angle, i.e. the angle between $\mu$ and $\pi$ momenta in the $J/\psi$ rest frame;
- $\tilde{\varphi}$ is the angle between the $(\mu^+, \mu^-)$ and $(K, \pi)$ planes in the $J/\psi$ rest frame;
- $\alpha$ is the angle between the $(\mu^+, \pi)$ and $(\mu^+, K\pi)$ planes in the $J/\psi$ rest frame.

The amplitudes for different $\lambda'$ values are related by parity conservation:

$$H_{\lambda'}^Z = -P(Z)(-1)^{J(Z)} H_{-\lambda'}^Z. \quad (11)$$

After inclusion of the Z component, the signal density function of Eq. (5) becomes,

$$S(\Phi) = \sum_{\xi=1,-1} \left| \sum_{K^*} \sum_{\lambda=-1,0,1} A_{\lambda\xi}^{K^*} + \sum_{Z} \sum_{\lambda'=-1,0,1} A_{\lambda'\xi}^Z \right|^2. \quad (12)$$

The signal density function of the charge conjugate decay, identified through the charge of the K (or $\pi$) differs only in the sign of $\varphi$. The implementation of this model takes into account this switching of sign and also allows for a possible flavour mis-tagging (typically a few %). For the full fit model with ten $K^*$s and two Zs as well as considering the floating masses and widths for some of the resonances, the total number of free parameters in the 4D probability density function can exceed 60. The large number of free parameters coupled with a complex PDF, which requires many internal mathematical operations to be executed at each step of the UML fit, poses a real computational challenge.

## Results

### Timing comparison

The computing capabilities of GPUs versus CPUs are tested by generating and fitting three sets, each comprising 10,000 Monte Carlo (MC) events (pseudo-experiments) of increasing complexity (number of $K^*$s) of the fit model previously described. The fitter implemented in ROOT/RooFit is run on an Intel Xeon cluster with 24 CPUs whereas the GooFit version is run on NVIDIA Tesla K40 GPU with 2880 CUDA cores. As the timing test models are for the demonstration purpose only, they are much less complex than the full model required for the analysis. Also, they process a smaller number of events than that expected from a collider experiment. As shown in Fig. 2, it becomes almost impossible to run the fitter on CPUs within any reasonable timescale when the number of fit parameters is increased. The GPU-based GooFit application provides a striking speed-up in performance compared to the CPU-based RooFit application. The latter gets so slow that it can become unreliable once the full number of parameters is adopted in the fit model.

### Fit validation

To validate the framework, a distribution according to the fit model is generated through MC techniques. These generated events mimic real data that are recorded by the collider experiments. A fit to that distribution is performed to check whether the best estimates of parameters returned by the fit are consistent with their input values.

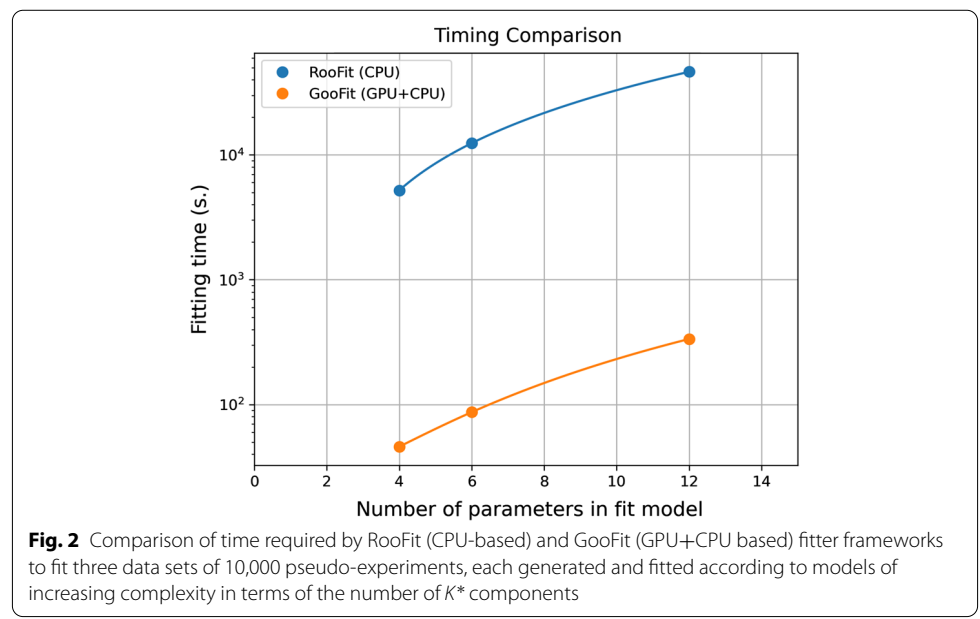

**Fig. 2** Comparison of time required by RooFit (CPU-based) and GooFit (GPU+CPU based) fitter frameworks to fit three data sets of 10,000 pseudo-experiments, each generated and fitted according to models of increasing complexity in terms of the number of $K^*$ components

### *Validation with the K\*-only model*

A pseudo-data sample of one million events is generated with the ten $K^*$s mentioned in "Methodology" section with their masses and widths fixed to the nominal values. The helicity amplitude parameters for each of these resonances are fixed to the values obtained by Belle [33].

As the PDF is four-dimensional, the fit results are presented as projections in each of the dimensions. The $m_{K\pi}$ projection of the fit to the generated dataset is shown in Fig. 3 and the other three projections, $m_{J/\psi\pi}$, $\cos\theta_{J/\psi}$, and $\varphi$, are presented in Fig. 4. The fit results are found to be in excellent agreement with the generated pseudo-data in each of the four dimensions signifying a good fit overall. The consistency of the post-fit values of the free parameters is checked by comparing the pull distributions (normalised residuals) with their generated values as shown in Figs. 5 and 6.

As the exact contribution of each resonance to the total signal cannot be precisely evaluated due to interference effects, an approximate measure is provided through the fit fractions. The fit fraction of the $j$-th resonance $R_j$ is given by:

$$FF_j = \frac{\int_\Omega |A^{R_j}(\Phi)|^2 d\mathbf{x}}{\int_\Omega |S(\Phi)|^2 d\mathbf{x}}, \qquad (13)$$

where $\Omega$ is the four-dimensional domain for the set of variables $\Phi$ [Eq. (1)] and $S(\Phi)$ is the signal function defined in Eq. (12). The numerator of Eq. (13) is obtained by setting to zero all the other helicity amplitudes at the post-fit level. The sum of all the fit fractions is not constrained to 100% as a consequence of the nonunitarity of the model which stems from the constructive and destructive effects of interference between the resonances.

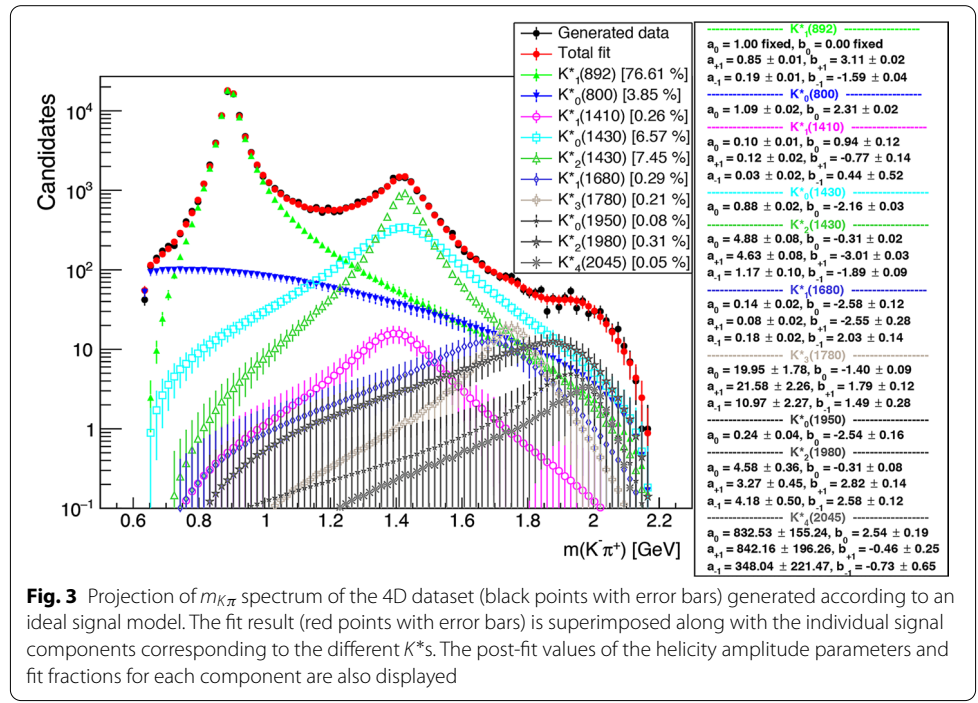

**Fig. 3** Projection of $m_{K\pi}$ spectrum of the 4D dataset (black points with error bars) generated according to an ideal signal model. The fit result (red points with error bars) is superimposed along with the individual signal components corresponding to the different $K^*$s. The post-fit values of the helicity amplitude parameters and fit fractions for each component are also displayed

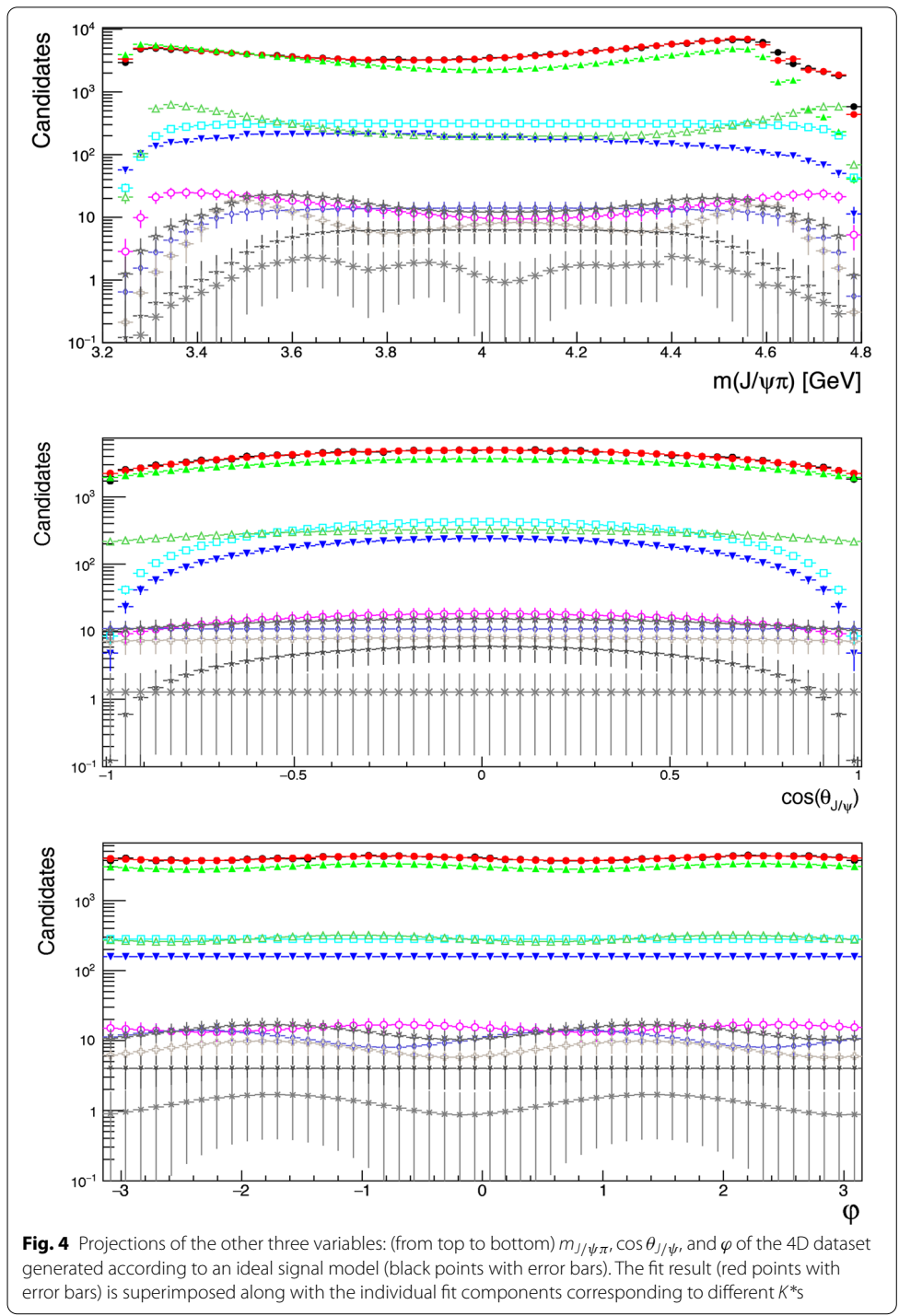

**Fig. 4** Projections of the other three variables: (from top to bottom) $m_{J/\psi\pi}$, $\cos\theta_{J/\psi}$, and $\varphi$ of the 4D dataset generated according to an ideal signal model (black points with error bars). The fit result (red points with error bars) is superimposed along with the individual fit components corresponding to different $K^*$s

### Sensitivity of the fitter to Z contributions

Fit validation exercises are performed for a) the $K^*$-only model but with the LASS lineshape used for the S-wave, and b) model with all ten $K^*$s together with $Z(4200)$ and $Z(4430)$ resonances. The mass, width, and helicity amplitudes of the $Z$ resonances are fixed to the values obtained by Belle [33]. It is found that the post-fit

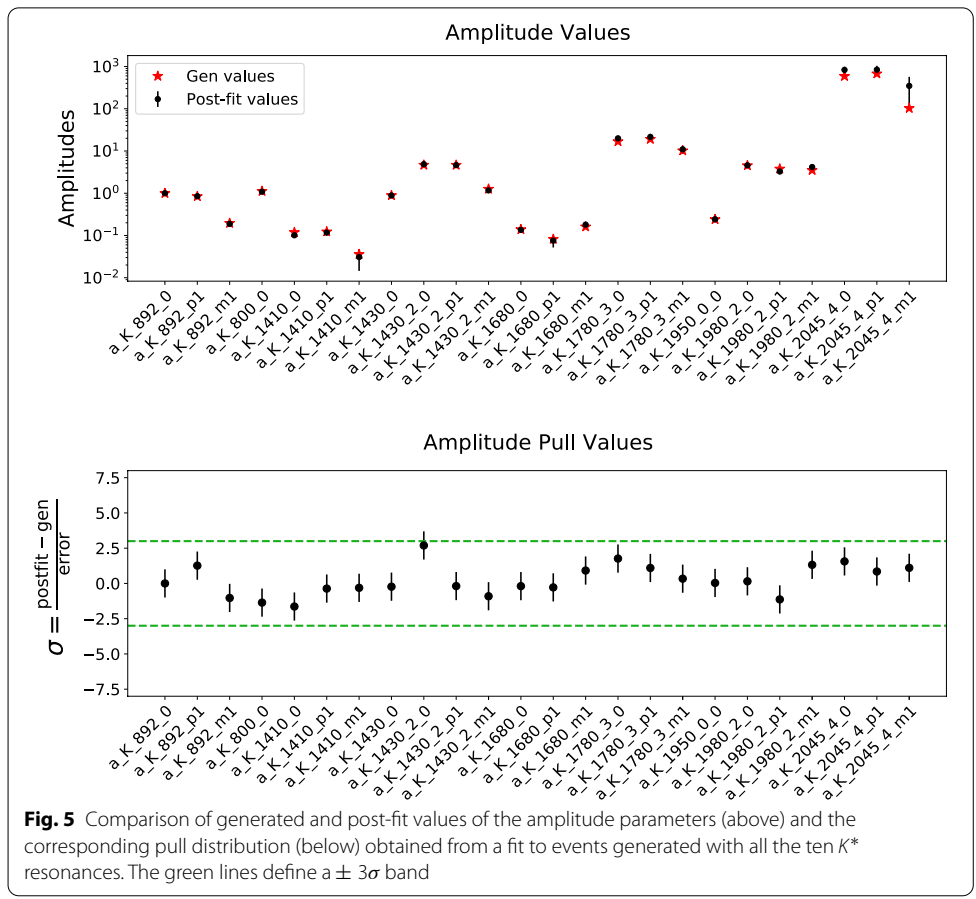

**Fig. 5** Comparison of generated and post-fit values of the amplitude parameters (above) and the corresponding pull distribution (below) obtained from a fit to events generated with all the ten $K^*$ resonances. The green lines define a $\pm 3\sigma$ band

values of parameters are consistent with the ones used for generation in both cases. The fit fractions of the $Z$-components are found to be small (about a few percent) as expected from the Belle results. This confirms that the fitter is capable of correctly detecting $Z$ contributions even if they are relatively small.

Since the $Z$ contributions are expected to be small, we need to ensure that the fitter does not artificially generate $Z$ peaks due to statistical fluctuations or alternative parametrizations of $K^*$ signals such as the LASS lineshape. Pseudo-data is generated with only ten $K^*$s and fitted with a [ten $K^*$s + $Z(4200)$ + $Z(4430)$] model. The fit fraction for both $Z(4200)$ and $Z(4430)$ are found to be 0.01%. From Figs. 7 and 8, it can be seen that the post-fit helicity amplitude values for the $K^*$s are close to their generated values indicating that the contribution of the $Z$s are indeed consistent with zero.

Similarly, another set of pseudo-data was generated with all $K^*$s (with LASS for the S-wave) and fitted with an "all $K^*$s (with LASS) + $Z(4200)$ + $Z(4430)$" model. The fit fractions for $Z(4200)$ and $Z(4430)$ are found to be 0.002% and 0.003%, respectively. Similar to the previous test, the post-fit helicity amplitude values for the $K^*$s are found to be close to their generated values signifying that the contribution of the $Z$s are again consistent with zero.

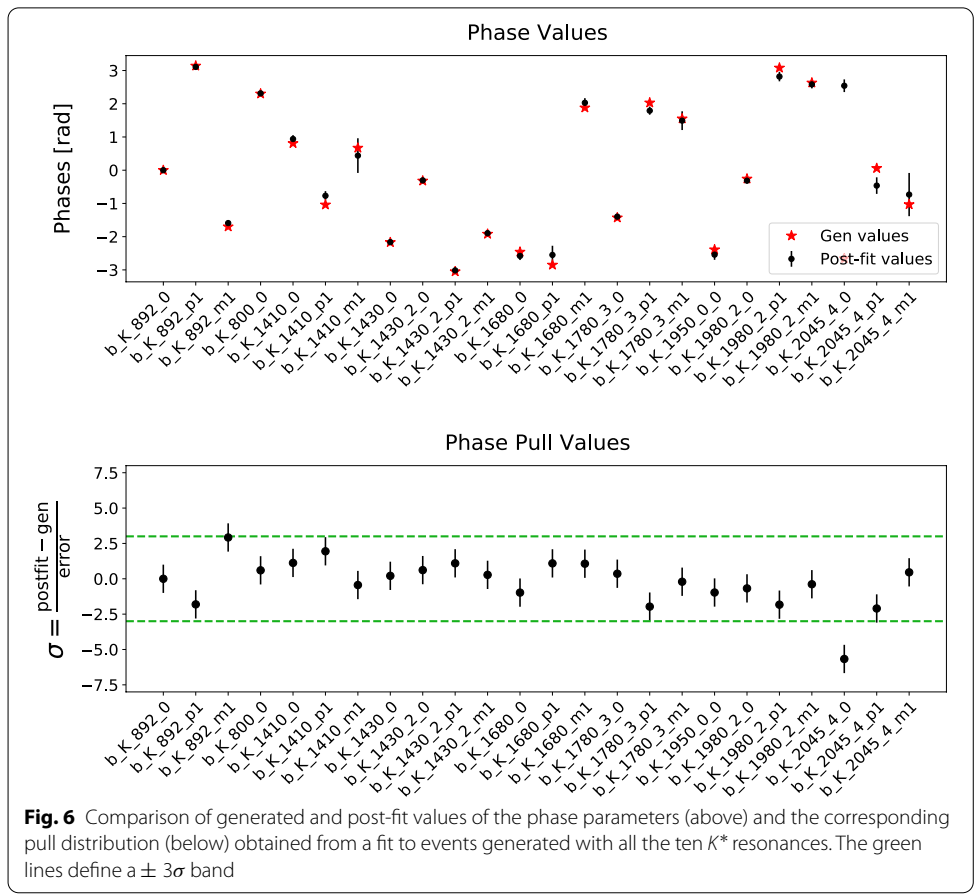

**Fig. 6** Comparison of generated and post-fit values of the phase parameters (above) and the corresponding pull distribution (below) obtained from a fit to events generated with all the ten $K^*$ resonances. The green lines define a $\pm 3\sigma$ band

**Applicability to real-life use cases**

An accurate representation of real data from collider experiments would require the inclusion of detection efficiency and background contamination. Keeping that in mind, the fit framework is developed in such a way that the efficiency and background models of suitable dimensions can be easily included in the form of analytical functions or binned templates. Generic shapes for efficiency (Fig. 9) and background (Fig. 10) in the form of 2D Bernstein polynomials are adopted to test the effectiveness of the fitter with efficiency and background included. Each of the 4D efficiency and background shapes is passed into the fitter as 2D (mass variables) × 2D (angular variables) histograms since the masses and angles are expected to be fully (or largely) uncorrelated.

Typically the background levels found in dedicated flavour-physics experiments (e.g. Belle and LHCb) are of the order of a few percent [33]. For this test, the fraction is set to a higher value keeping in mind general purpose detectors like CMS and ATLAS that may record signals with less purity due to the absence of dedicated hadron identification systems. One million simulated decays are generated and fitted with a model including all ten $K^*$s, two $Z$s as well as the relative efficiency and background parametrizations. The relative efficiencies are used to weight the signal model, whereas the background is added with a fixed coefficient (equal to [1—signal purity]

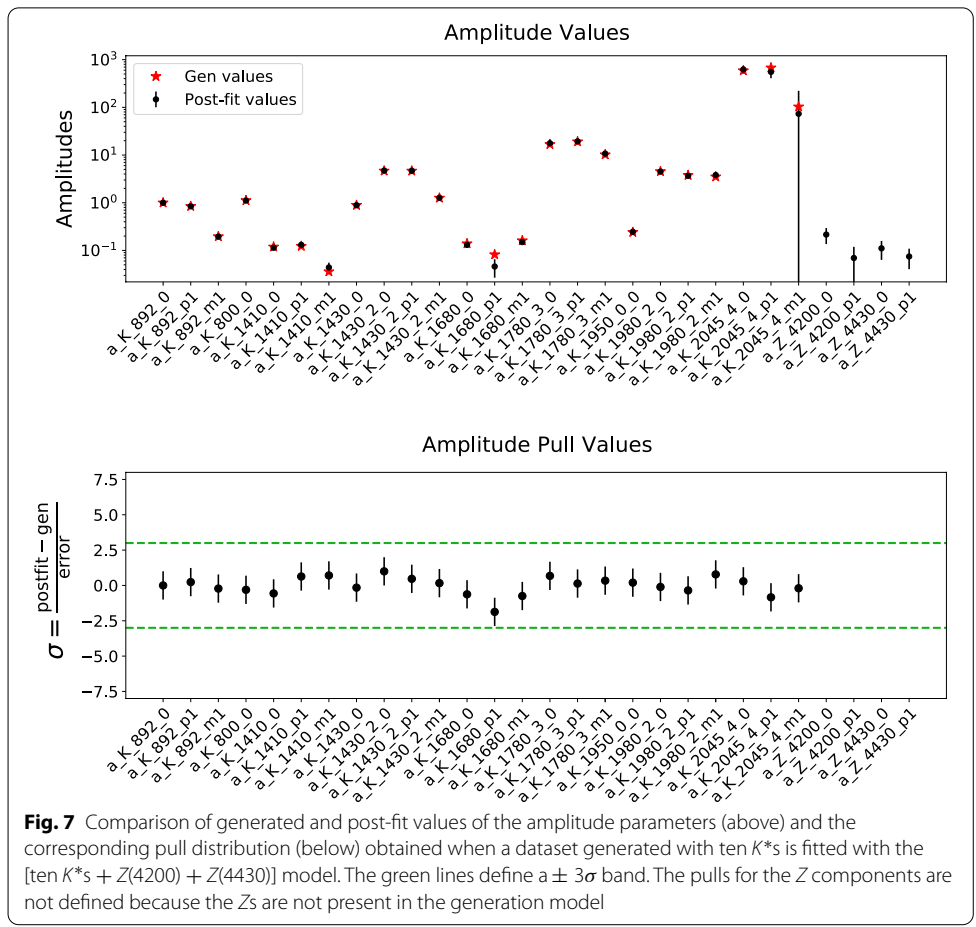

**Fig. 7** Comparison of generated and post-fit values of the amplitude parameters (above) and the corresponding pull distribution (below) obtained when a dataset generated with ten $K^*$s is fitted with the [ten $K^*$s + $Z(4200)$ + $Z(4430)$] model. The green lines define a $\pm 3\sigma$ band. The pulls for the $Z$ components are not defined because the $Z$s are not present in the generation model

which in this study is assumed to be 15%). Therefore, the 4D PDF $f(\Phi)$, on which the UML fit is to be performed, takes the form:

$$f(\Phi) = p \cdot \epsilon(\Phi) \cdot S(\Phi) + (1-p) \cdot b(\Phi), \tag{14}$$

where

- $p$ is the signal purity;
- $\epsilon(\Phi)$ is the 4D relative signal efficiency;
- $S(\Phi)$ is the signal density function defined in Eq. (12) and
- $b(\Phi)$ is the 4D background PDF model.

From Figs. 11 and 12, it can be seen that the post-fit helicity amplitude values for the $K^*$s and the $Z$s are close to their generated values. The fit fractions of $Z(4200)$ (3.49%) and $Z(4200)$ (1.17%) are found to be a few percent as expected from the Belle result [33]. The almost identical post-fit values of the parameters from both the ideal-world case and the real-life case indicate that the fitter can produce reliable results while taking into account the detection efficiency and background contributions.

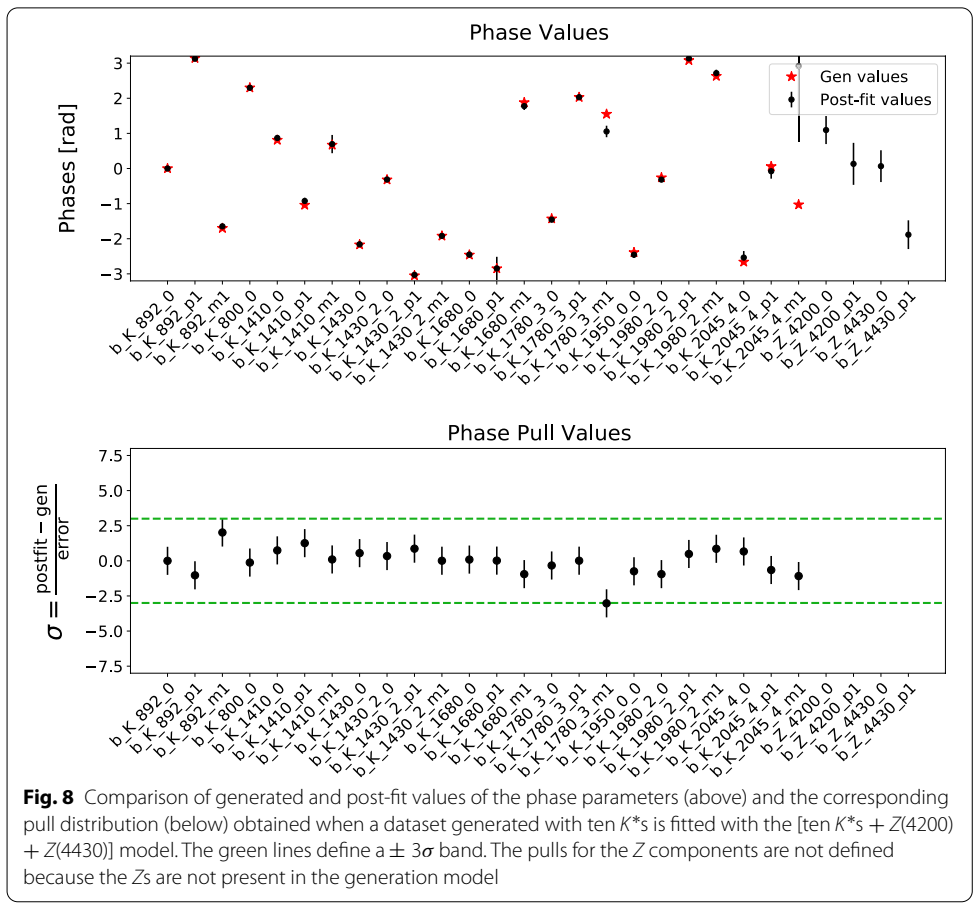

**Fig. 8** Comparison of generated and post-fit values of the phase parameters (above) and the corresponding pull distribution (below) obtained when a dataset generated with ten $K^*$s is fitted with the [ten $K^*$s + $Z$(4200) + $Z$(4430)] model. The green lines define a $\pm 3\sigma$ band. The pulls for the $Z$ components are not defined because the $Z$s are not present in the generation model

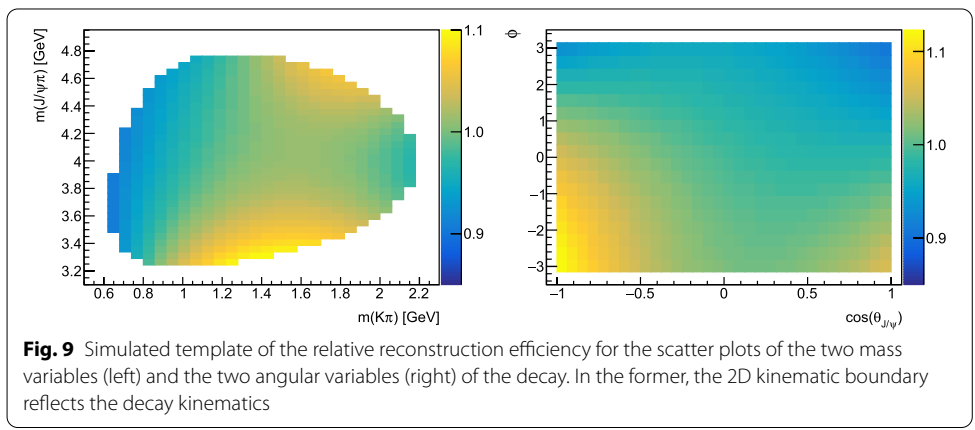

**Fig. 9** Simulated template of the relative reconstruction efficiency for the scatter plots of the two mass variables (left) and the two angular variables (right) of the decay. In the former, the 2D kinematic boundary reflects the decay kinematics

## Discussion

Searches for exotic multiquark states in collider experiments require complex multidimensional analyses involving several (or even hundreds of) thousands of events that demand considerable computational resources. Conventional CPU-based techniques may fall short to meet these ever increasing demands. In this study, by using the helicity formalism, a four-dimensional amplitude analysis framework for an unbinned maximum-likelihood fit has been implemented. The fitting framework has been developed using the novel GPU based GooFit, an open-source tool under development which is used in HEP applications for parameters estimation, interfacing ROOT to the CUDA parallel computing platform on NVIDIA GPUs. It has been shown that the choice to use GooFit and the accelerated performance provided by GPUs is crucial to carry out these extreme fits.

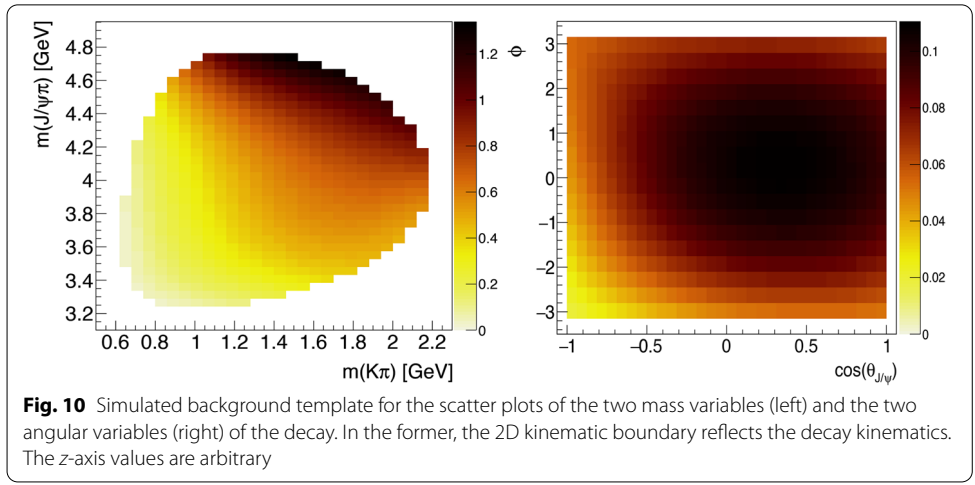

**Fig. 10** Simulated background template for the scatter plots of the two mass variables (left) and the two angular variables (right) of the decay. In the former, the 2D kinematic boundary reflects the decay kinematics. The *z*-axis values are arbitrary

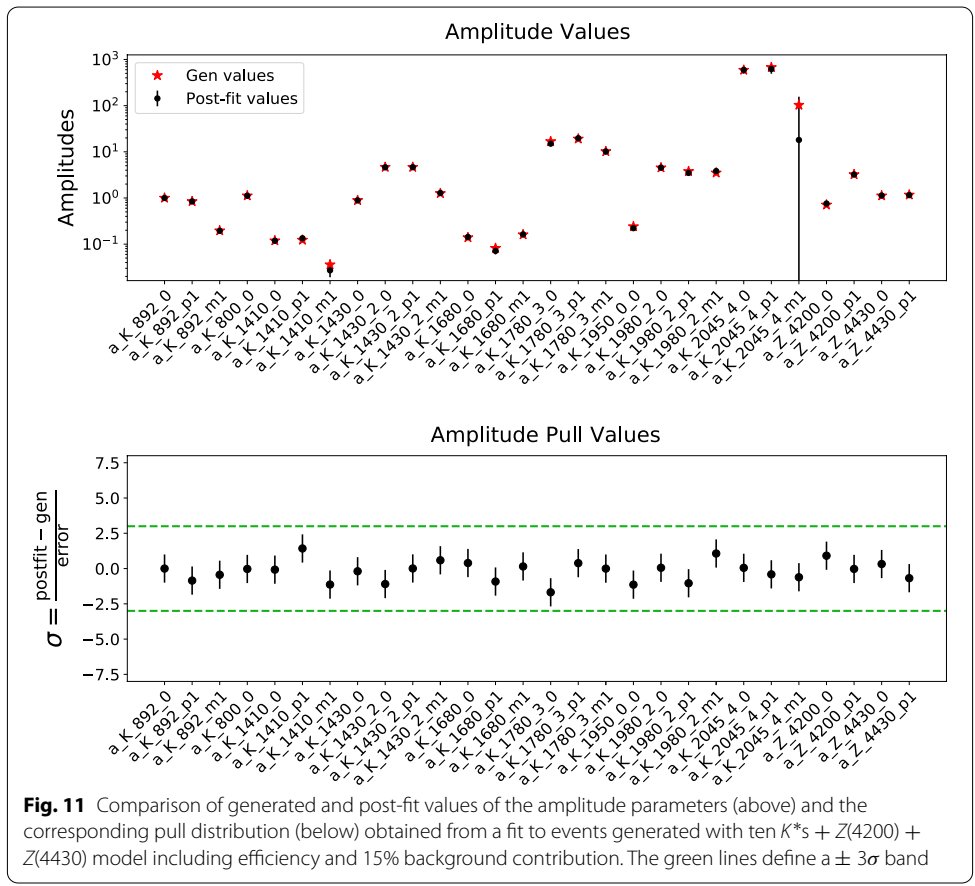

**Fig. 11** Comparison of generated and post-fit values of the amplitude parameters (above) and the corresponding pull distribution (below) obtained from a fit to events generated with ten $K^*$s + $Z(4200)$ + $Z(4430)$ model including efficiency and 15% background contribution. The green lines define a $\pm 3\sigma$ band

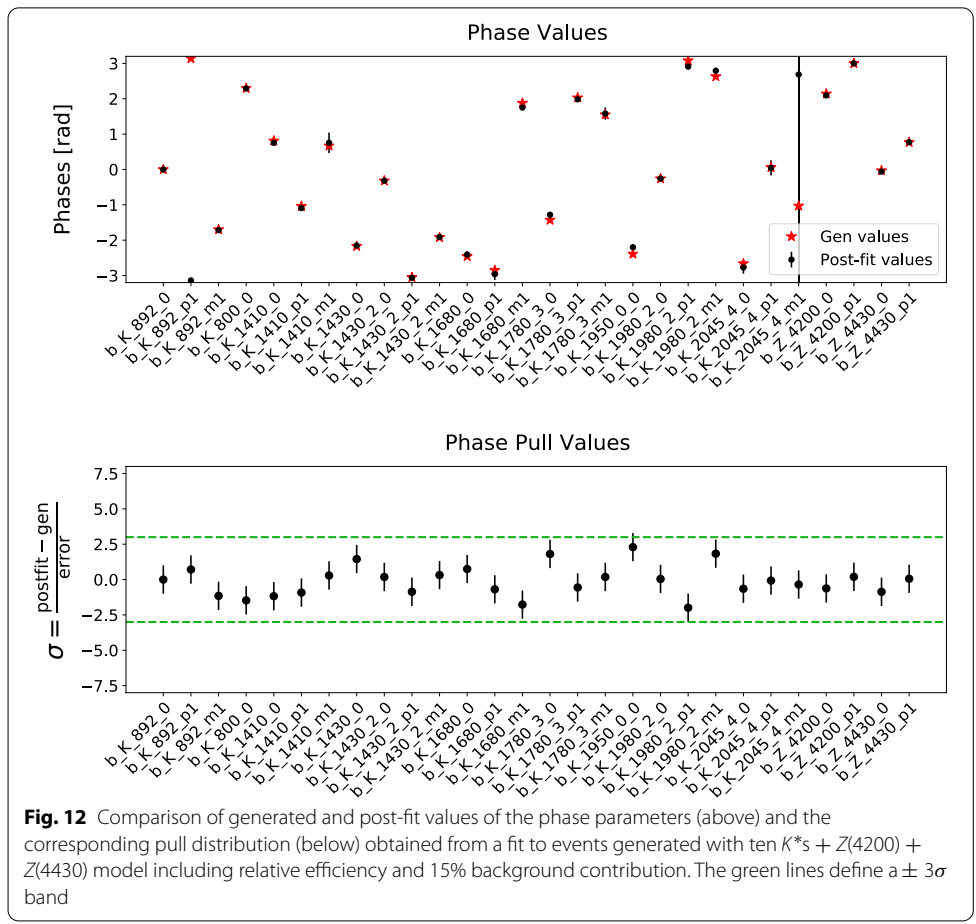

**Fig. 12** Comparison of generated and post-fit values of the phase parameters (above) and the corresponding pull distribution (below) obtained from a fit to events generated with ten $K^*$s + $Z(4200)$ + $Z(4430)$ model including relative efficiency and 15% background contribution. The green lines define a $\pm\,3\sigma$ band

The fit model has been validated by a "closure test", i.e., by a multi-step procedure in which pseudo-experiments under different conditions and assumptions were generated and fitted. The starting model is assumed to be composed of the known set of $K^*$ resonances. Since the low mass S-wave $K_0^*(800)$ is not yet satisfactorily described by a Breit–Wigner amplitude, the alternative LASS parametrization has been implemented on GooFit and thoroughly tested. The fitter has been additionally equipped with the capability of handling relative detection efficiency and background contamination. The possible contribution of the exotic $Z$ states has been calculated and incorporated within the fitter framework with reasonable robustness to allow for testing any combination of their spin-parity values as well as without any constraints. Lastly, the fitter, though designed for a 4D amplitude analysis of a pseudoscalar decaying into a vector and two pseudoscalars, can be easily adapted to other types of decays with higher or lower dimensions, occurring in flavour physics studies.

## Conclusion

The ability of the fitter to efficiently handle higher dimensionality of fit models with great accuracy, its inbuilt functions to calculate complex operations like vector algebra while evaluating PDFs on the GPU-side, its systematized application of Gaussian constraints on fit parameters if required, and its sensitivity to very small contributions of different varieties of hitherto unknown signals make it a formidable toolkit built into an already powerful framework. It is hoped that this kind of fitter implemented within the GooFit framework, along with the flexibility to be easily adapted for even more complex PDFs, will considerably augment the capabilities of collider experiments in searches and measurements in the field of exotic hadron spectroscopy and beyond.

**Abbreviations**
GPU: Graphics processing unit; QCD: Quantum chromodynamics; UML: Unbinned maximum likelihood; HEP: High-energy physics; BW: Breit–Wigner; PDF: Probability density function; NLL: Negative log likelihood; MC: Monte Carlo.

**Acknowledgements**
We acknowledge Prof. Alexis Pompili (Universitá degli Studi di Bari and I.N.F.N. - Sezione di Bari) and Prof. Gagan Mohanty (Tata Institute of Fundamental Research, Mumbai) for their guidance and constant support as well as helpful comments on this article.
   We are indebted to Prof. John Yelton (University of Florida) for his valuable comments and suggestions on the content as well as the language and presentation of this article.
   The computational work has been executed on the IT resources of the ReCaS-Bari data centre, which have been made available by two projects financed by the MIUR (Italian Ministry for Education, University and Research) in the "PON Ricerca e Competitività 2007-2013" Program: ReCaS (Azione I - Interventi di rafforzamento strutturale, PONa3_00052, Avviso 254/Ric) and PRISMA (Asse II - Sostegno all'innovazione, PON04a2_A).

**Authors' contributions**
All authors have equal contribution. All authors read and approved the final manuscript.

**Funding**
Open access funding is partially provided by Tata Institute of Fundamental Research, Mumbai, India and Universitá degli Studi di Bari and I.N.F.N. - Sezione di Bari, Italy. The research received no external funding.

**Availability of data and materials**
The datasets generated and analysed during the current study are available from the corresponding author on reasonable request.

**Ethics approval and consent to participate**
Not applicable.

**Consent for publication**
Not applicable.

**Competing interests**
The authors declare that they have no competing interests.

**Author details**
[1] Tata Institute of Fundamental Research, Mumbai, India. [2] Universitá degli Studi di Bari and I.N.F.N. - Sezione di Bari, Bari, Italy.

Received: 31 July 2020   Accepted: 27 December 2020
Published online: 07 January 2021

## Publisher's Note